\documentclass[aps,pra,twocolumn,showpacs]{revtex4}
\usepackage{amsfonts}
\usepackage{amsmath}
\usepackage{amssymb}
\usepackage{graphicx}
\usepackage[english]{babel}
\usepackage{hyperref}
\usepackage[usenames,dvipsnames]{color}
\usepackage{babel}

\begin{document}

\title{Two-photon exchange interaction from Dicke Hamiltonian under
parametric modulation}
\author{A. V. Dodonov }
\email{adodonov@fis.unb.br}
\affiliation{Institute of Physics and International Centre for Condensed Matter Physics,
University of Brasilia, 70910-900, Brasilia, Federal District, Brazil}

\begin{abstract}
We consider the nonstationary circuit QED architecture in which a single-mode
cavity interacts with $N>1$ identical qubits, and some
system parameters undergo a weak external perturbation. It is shown that in
the dispersive regime one can engineer the two-photon exchange interaction
by adjusting the frequency of harmonic modulation to (approximately) $%
2|\Delta _{-}|$, where $\Delta _{-}$ is the average atom--field detuning.
Closed analytic description is derived for the weak atom--field coupling
regime, and numeric simulations indicate that the phenomenon can be
observed in the present setups.
\end{abstract}

\pacs{42.50.Pq, 42.50.Ct, 42.50.Hz, 32.80-t, 03.65.Yz}
\keywords{nonstationary circuit QED, Two-photon exchange interaction,
Tavis-Cummings Hamiltonian, Dicke Hamiltonian}
\maketitle

\section{Introduction}

The area of circuit Quantum Electrodynamics (circuit QED) has grown to
embrace a plethora of architectures with different kinds of multi-level
atoms and sophisticated assemblies of interconnected 3D and 1D resonators
and waveguides \cite{paik,you,science,rev1,APL,rev2,nori2017}. Diverse
designs incorporating up to tens of Josephson Junctions give rise to
superconducting artificial atoms with distinct properties regarding the
dissipation mechanisms and the structure of energy levels, however, all they
share the capability of coherently coupling to the Electromagnetic (EM)\
field \cite{nor,smith,reagor,wilson,ofec,wallr}. Moreover, a single cavity
mode can interact with several locally addressable artificial atoms \cite{3at0,3at1,3at2,ann,cle,saito,gam1,song} or an ensemble of trapped ultracold
atoms \cite{uatoms1,uatoms2}.

Many types of superconducting artificial atoms allow for real-time manipulation of the
energy levels or the atom--field coupling strength \cite{majer,ge,ger,ger1,v1,v2,v3}. Combined with the ability of \emph{in situ}
tuning the resonator's frequency by external magnetic flux \cite{nori-n,meta}, such nonstationary circuit QED architectures give rise to a
novel regime of light--matter interaction in which all the parameters in the
Hamiltonian are controllable functions of time \cite{jpcs,JPA}.
Using resonant perturbations one can induce creation and annihilation of
photons or atomic excitations \cite{liberato9,roberto,diego,etc1,juan,hoeb},
generate entanglement \cite{entangles,etc3,enta2}, induce new forms of
light--matter interaction \cite{igor,porras,palermo,tom}, perform quantum
simulations \cite{relativistic,sim} and study other novel effects \cite{etc2,processing,ert,ert2}. Some of the
early proposals \cite{jpcs,bla} have recently been verified experimentally, such as the
one-photon exchange between the qubit and the field in the dispersive regime
(reliant on the \lq rotating\rq\ terms in the
interaction Hamiltonian) \cite{blais-exp,simmonds} and generation of two
quanta from vacuum due to the \lq counter-rotating
terms\rq\ (CRT) \cite{schuster}.

In this paper we describe another effect based on the rotating\ terms -- the
\emph{two-photon exchange interaction} -- that can be implemented in
nonstationary circuit QED by modulating any system parameter with frequency $%
\eta \approx 2|\Delta _{-}|$, where $\Delta _{-}$ is the average atom--field
detuning. We illustrate the phenomenon for the case of $N>1$ off-resonant
qubits described by the Dicke \cite{dicke,HP} or Tavis-Cummings \cite{TC1,TC2} Hamiltonians, however, our approach can be straightforwardly
generalized to an arbitrary multilevel atom in the ladder configuration \cite{JPA}. Assuming the weak atom--field coupling regime, we derive a closed
analytic description of the unitary dynamics (see Sec. \ref{disp}) and find
a good agreement with numeric data even for moderate coupling strengths
(Sec. \ref{sec3}). We also show that our proposal can be implemented in
the current circuit QED setups with weak dissipation and slightly different
atoms (Sec. \ref{seccc}), and discuss manners to enhance the two-photon transition rate.

\section{Analytic results}

\label{disp}

Our system consists of a single mode of EM field\ interacting with $N$
qubits, as described by the {\em quantum Dicke model} \cite{dicke,HP}%
\begin{equation}
\hat{H}/\hbar =\omega \hat{n}+\sum_{l=1}^{N}\left[ \frac{\Omega }{2}\hat{%
\sigma}_{z}^{(l)}+g(\hat{a}+\hat{a}^{\dagger })(\hat{\sigma}_{+}^{(l)}+\hat{%
\sigma}_{-}^{(l)})\right] ~,  \label{H1}
\end{equation}%
where the index $l$ labels the identical noninteracting atoms. We assume
that the cavity frequency $\omega $, the atomic transition frequency $\Omega
$ and the atom--cavity coupling strength$~g$ are externally prescribed
functions of time ($g$ is considered real). $\hat{a}$ and $\hat{a}%
^{\dagger }$ are the annihilation and creation operators and $\hat{n}=\hat{a}%
^{\dagger }\hat{a}$ is the photon number operator. The qubit operators are $%
\hat{\sigma}_{-}^{(l)}=|g^{(l)}\rangle \langle e^{(l)}|$, $\hat{\sigma}%
_{+}^{(l)}=|e^{(l)}\rangle \langle g^{(l)}|$ and $\hat{\sigma}%
_{z}^{(l)}=|e^{(l)}\rangle \langle e^{(l)}|-|g^{(l)}\rangle \langle g^{(l)}|$%
, where $|g^{(l)}\rangle $ and $|e^{(l)}\rangle $ denote the\ ground and
excited states of the $l$-th qubit, respectively. In the absence of CRT $%
\sum_{l=1}^{N}(\hat{a}\hat{\sigma}_{-}^{(l)}+\hat{a}^{\dagger }\hat{\sigma}%
_{+}^{(l)})$ the Hamiltonian (\ref{H1}) is known as \emph{Tavis-Cummings
Hamiltonian} \cite{TC1,TC2}. We stress that although our approach takes into
account the CRT, the phenomenon described in this paper does not
require their presence.

We consider the general case of simultaneous external modulation of all the system parameters
as $X=X_{0}+\varepsilon _{X}\sin (\eta t+\phi _{X})$, where $X$ stands for $%
\omega $, $\Omega $ or $g$ (for the particular case of a single-parameter modulation, only one $\varepsilon _{X}$ is nonzero). $X_{0}$ is the bare value, $\varepsilon _{X}\geq
0$ is the respective modulation depth, $\eta \sim 2|\omega _{0}-\Omega _{0}|$
is the modulation frequency and $\phi _{X}$ is the initial phase. Moreover,
we restrict our analysis to the perturbative regime when $\varepsilon
_{g}\ll g_{0}$ and $\varepsilon _{\omega },\varepsilon _{\Omega }\ll |\omega
_{0}-\Omega _{0}|$.

To obtain closed analytical description we employ the normalized \emph{Dicke}
states with $k$ atomic excitations%
\begin{equation}
|\mathbf{k}\rangle =\sqrt{\frac{k!(N-k)!}{N!}}\sum_{p}|e^{\left( 1\right)
}\rangle |e^{\left( 2\right) }\rangle \cdots |e^{\left( k\right) }\rangle
|g^{\left( k+1\right) }\rangle \cdots |g^{N}\rangle ,
\end{equation}%
where the sum runs over all the allowed permutations of excited and
non-excited qubits and $k=0,1,\ldots ,N$. In the Dicke basis the Hamiltonian
(\ref{H1}) reads%
\begin{equation}
\hat{H}/\hbar =\omega \hat{n}+\sum_{k=0}^{N}[\Omega k\hat{\sigma}%
_{k,k}+gf_{k}(\hat{a}+\hat{a}^{\dagger })(\hat{\sigma}_{k+1,k}+\hat{\sigma}%
_{k,k+1})],  \label{bsb}
\end{equation}%
where $\hat{\sigma}_{k,j}\equiv |\mathbf{k}\rangle \langle \mathbf{j}|$ and $%
f_{k}\equiv \sqrt{(k+1)(N-k)}$.

We work in the \emph{dispersive regime}, $g_{0}f_{k}\sqrt{n}\ll
|\Delta _{-}|$ for all relevant values of $n$, where $\Delta _{-}=\omega
_{0}-\Omega _{0}$ is the bare atom--field detuning and $n$ is the total
number of excitations. Following the approach described in \cite{JPA,juan},
we expand the system state as%
\begin{equation}
|\psi (t)\rangle =\sum_{m=0}^{\infty }\sum_{\mathcal{S}}e^{i\Phi _{m,%
\mathcal{S}}(t)}e^{-it\tilde{\lambda}_{m,\mathcal{S}}}b_{m,\mathcal{S}%
}(t)|\varphi _{m,\mathcal{S}}\rangle ~.  \label{psit}
\end{equation}%
$\Phi _{m,\mathcal{S}}(t)$ is a real oscillatory function%
\begin{equation}
\Phi _{m,\mathcal{S}}(t)=\sum_{k=0}^{N}\sum_{L=\omega ,g,\Omega }\frac{%
\Upsilon _{m,\mathcal{S},\mathcal{S}}^{L,k}}{\eta }\left[ \cos (\eta t+\phi
_{L})-\cos \phi _{L}\right] \,,~
\end{equation}%
where we defined the constant coefficients%
\begin{equation}
\Upsilon _{m,\mathcal{T},\mathcal{S}}^{\omega ,k}\equiv \delta
_{k,0}\varepsilon _{\omega }\langle \varphi _{m,\mathcal{T}}|\hat{n}|\varphi
_{m,\mathcal{S}}\rangle
\end{equation}%
\begin{equation}
\Upsilon _{m,\mathcal{T},\mathcal{S}}^{g,k}\equiv \varepsilon
_{g}f_{k}\langle \varphi _{m,\mathcal{T}}|(\hat{a}\hat{\sigma}_{k+1,k}+\hat{a%
}^{\dagger }\hat{\sigma}_{k,k+1})|\varphi _{m,\mathcal{S}}\rangle
\end{equation}%
\begin{equation}
\Upsilon _{m,\mathcal{T},\mathcal{S}}^{\Omega ,k}\equiv \varepsilon _{\Omega
}k\langle \varphi _{m,\mathcal{T}}|\hat{\sigma}_{k,k}|\varphi _{m,\mathcal{S}%
}\rangle
\end{equation}%
and assumed that $|\Upsilon _{n,\mathcal{T},\mathcal{S}}^{L,k}/\Delta
_{-}|\ll 1$ for $L=\omega ,g,\Omega $ and all relevant values of $k,n$, $%
\mathcal{T}$ and $\mathcal{S}$.

$\tilde{\lambda}_{m,\mathcal{S}}\equiv \lambda _{m,\mathcal{S}}+\nu _{m,%
\mathcal{S}}$ is the effective eigenfrequency, where $\lambda _{m,\mathcal{S}%
}$ and $|\varphi _{m,\mathcal{S}}\rangle $ are the $m$-excitations
eigenfrequencies and eigenstates (\emph{dressed states}) of the bare
Hamiltonian%
\begin{equation}
\hat{H}_{0}/\hbar =\omega _{0}\hat{n}+\sum_{k=0}^{N}[\Omega _{0}k\hat{\sigma}%
_{k,k}+g_{0}f_{k}(\hat{a}\hat{\sigma}_{k+1,k}+\hat{a}^{\dagger }\hat{\sigma}%
_{k,k+1})]\,.
\end{equation}%
The index $\mathcal{S}$ labels the different eigenvalues and eigenstates
within the subspace of a given value of $m$. $\nu _{m,\mathcal{S}}\sim
\mathcal{O}(g_{0}^{2}/\omega _{0})$ is the frequency shift \cite{juan} due
to the counter-rotating terms in Eq. (\ref{bsb}):%
\begin{eqnarray}
\nu _{m,\mathcal{T}} &=&g_{0}^{2}\sum_{\mathcal{S}}\left[ \frac{\left(
\sum_{k=0}^{N}f_{k}\Lambda _{k,m,\mathcal{S},\mathcal{T}}\right) ^{2}}{%
\lambda _{m,\mathcal{T}}-\lambda _{m-2,\mathcal{S}}}\right.  \notag \\
&&\left. -\frac{\left( \sum_{k=0}^{N}f_{k}\Lambda _{k,m+2,\mathcal{T},%
\mathcal{S}}\right) ^{2}}{\lambda _{m+2,\mathcal{S}}-\lambda _{m,\mathcal{T}}%
}\right]
\end{eqnarray}%
\begin{equation}
\Lambda _{k,m+2,\mathcal{T},\mathcal{S}}\equiv \langle \varphi _{m,\mathcal{T%
}}|\hat{a}\hat{\sigma}_{k,k+1}|\varphi _{m+2,\mathcal{S}}\rangle \,,
\end{equation}%
where we assumed $g_{0}f_{k}\Lambda _{k,m+2,\mathcal{S},\mathcal{T}}\lesssim
\omega _{0}-\left\vert \Delta _{-}\right\vert $. Lastly, $b_{m,\mathcal{S}%
}(t)$ is the {\em slowly-varying probability amplitude} for the dressed state $%
|\varphi _{m,\mathcal{S}}\rangle $.

Substituting (\ref{psit}) into the Schr\"{o}dinger equation and neglecting
the rapidly oscillating terms \cite{JPA},\ we obtain to the first order in $%
\varepsilon _{\omega }$, $\varepsilon _{g}$ and $\varepsilon _{\Omega }$%
\begin{equation}
\dot{b}_{n,\mathcal{T}}=\sum_{\mathcal{S}\neq \mathcal{T}}\Xi _{n,\mathcal{T}%
,\mathcal{S}}e^{its_{n,\mathcal{T},\mathcal{S}}(|\tilde{\lambda}_{n,\mathcal{%
T}}-\tilde{\lambda}_{n,\mathcal{S}}|-\eta )}b_{n,\mathcal{S}}~,  \label{bor}
\end{equation}%
where $s_{n,\mathcal{T},\mathcal{S}}\equiv \mathrm{sign}(\tilde{\lambda}_{n,%
\mathcal{T}}-\tilde{\lambda}_{n,\mathcal{S}})$ and%
\begin{equation}
\Xi _{n,\mathcal{T},\mathcal{S}}\equiv \frac{s_{n,\mathcal{T},\mathcal{S}}}{2%
}\sum_{k=0}^{N}\sum_{L=g,\Omega ,\omega }\Upsilon _{n,\mathcal{T},\mathcal{S}%
}^{L,k}e^{-is_{n,\mathcal{T},\mathcal{S}}\phi _{L}}~,  \label{dor}
\end{equation}%
$\Xi _{n,\mathcal{T},\mathcal{S}}^{\ast }=-\Xi _{n,\mathcal{S},\mathcal{T}}$%
. To obtain Eq. (\ref{bor}) we neglected frequency shifts smaller than $\nu
_{n,\mathcal{T}}$, as well as the corrections of the order of $(\Upsilon _{n,%
\mathcal{T},\mathcal{S}}^{L,k})^{2}/\Delta _{-}\propto \varepsilon
_{L}^{2}/\Delta _{-}$ for $L=\omega ,g,\Omega $.

Under the resonant modulation frequency, $\eta _{res}=|\tilde{\lambda}_{n,%
\mathcal{T}}-\tilde{\lambda}_{n,\mathcal{S}}|$, we get
\begin{equation}
b_{n,\mathcal{T}}=b_{n,\mathcal{T}}\left( 0\right) \cos |\Xi _{n,\mathcal{T},%
\mathcal{S}}|t+\frac{\Xi _{n,\mathcal{T},\mathcal{S}}}{|\Xi
_{n,\mathcal{T},\mathcal{S}}|}b_{n,\mathcal{S}}\left( 0\right) \sin |\Xi _{n,\mathcal{T},%
\mathcal{S}}|t \, ,  \label{gor}
\end{equation}%
corresponding to the modulation-induced transition $|\varphi _{n,\mathcal{T}%
}\rangle \leftrightarrow |\varphi _{n,\mathcal{S}}\rangle $. We emphasize that,
under the considered approximations, Eqs. (\ref{bor}) -- (\ref{dor}) also
describe the effective dynamics in the absence of CRT in the Hamiltonian (%
\ref{bsb}), though in this case the frequency shift $\nu _{m,\mathcal{T}}=0$
so that $\tilde{\lambda}_{n,\mathcal{T}}=\lambda _{n,\mathcal{T}}$ .

In the dispersive regime the spectrum of $\hat{H}_{0}$ can be obtained from
the standard perturbation theory. To the second order in $g_{0}/\Delta _{-}$
one finds%
\begin{equation*}
\lambda _{n,k}=n\omega _{0}-k\Delta _{-}+\delta _{-}\left[
(N-k)(n-2k)-k(n-k+1)\right]
\end{equation*}%
\begin{eqnarray}
|\varphi _{n,k}\rangle &=&\mathcal{N}_{n,k}\left[ |\varphi _{n,k}^{\left(
0\right) }\rangle +\frac{g_{0}f_{k}\sqrt{K}}{\Delta _{-}}|\varphi
_{n,k+1}^{\left( 0\right) }\rangle \right.  \label{dress} \\
&&-\frac{g_{0}f_{k-1}\sqrt{K+1}}{\Delta _{-}}|\varphi _{n,k-1}^{\left(
0\right) }\rangle  \notag \\
&&+\frac{g_{0}^{2}f_{k}f_{k+1}\sqrt{K\left( K-1\right) }}{2\Delta _{-}^{2}}%
|\varphi _{n,k+2}^{\left( 0\right) }\rangle  \notag \\
&&\left. +\frac{g_{0}^{2}f_{k-1}f_{k-2}\sqrt{\left( K+1\right) \left(
K+2\right) }}{2\Delta _{-}^{2}}|\varphi _{n,k-2}^{\left( 0\right) }\rangle %
\right] ,  \notag
\end{eqnarray}%
where $k=0,1,2,...,\min \left( n,N\right) $, $K=n-k$, $\delta
_{-}=g_{0}^{2}/\Delta _{-}$, $|\varphi _{n,k}^{\left( 0\right) }\rangle =|%
\mathbf{k},n-k\rangle \equiv |\mathbf{k}\rangle _{atom}\otimes |n-k\rangle
_{field}$ and $\mathcal{N}_{n,k}$ is the normalization constant ($|n\rangle
_{field}$ is the cavity Fock state). Actually, to evaluate the two-photon
transition rate one needs the eigenstates to the fourth order in $%
g_{0}/\Delta _{-}$, which are omitted here for brevity.

The {\em two-photon exchange interaction} \cite{chen,alexanian,chen2} corresponds
to the transition $|\varphi _{n,k}\rangle \leftrightarrow |\varphi
_{n,k+2}\rangle $, which represents (approximately) $|\mathbf{k},n-k\mathbf{%
\rangle }\leftrightarrow |\mathbf{k+2},n-k-2\rangle $. To the lowest order
in $g_{0}/\Delta _{-}$ we obtain%
\begin{eqnarray}
\Xi _{n,k,k+2} &= &\mathcal{D}g_{0}\left( \frac{g_{0}}{\Delta _{-}}%
\right) ^{3}\sqrt{(N-k)(N-k-1)}  \notag \\
&&\times \sqrt{(k+1)(k+2)K(K-1)}  \label{rate} \\
&&\times \left( \frac{\varepsilon _{\omega }e^{-i\mathcal{D}\phi _{\omega }}%
}{\Delta _{-}}-\frac{\varepsilon _{\Omega }e^{-i\mathcal{D}\phi _{\Omega }}}{%
\Delta _{-}}-\frac{\varepsilon _{g}e^{-i\mathcal{D}\phi _{g}}}{g_{0}}\right)
,  \notag
\end{eqnarray}%
where $\mathcal{D}=\mathrm{sign}(\Delta _{-})$. The corresponding resonant
modulation frequency reads%
\begin{equation}
\eta _{r}\approx 2\left\vert \Delta _{-}+\delta _{-}(2N+2n-6k-5)\right\vert
\,.  \label{lor}
\end{equation}%
Notice that for a given value of $k$ the other states $\{|\mathbf{k}%
,n^{\prime }-k\mathbf{\rangle ,}|\mathbf{k+2},n^{\prime }-k-2\rangle \}$ ($%
n^{\prime }\neq n$) are not affected by such modulation due to the condition
$|\delta _{-}|\gg |\Xi _{n,k,k+2}|$, as can be seen from Eq. (\ref{bor}). So
in the most common situation when the atoms are initially in the ground
states (the Dicke state $|\mathbf{0}\rangle $) only the pair of states
selected by the modulation frequency becomes coupled. On
the other hand, if the atoms are prepared in a superposition of Dicke
states, several states can become coupled by a single modulation frequency.
For example, from Eq. (\ref{lor}) we see that to the second order in $%
g_{0}/\Delta _{-}$ the same frequency $\eta _{r}$ couples the pair of states
$|\mathbf{0},4\mathbf{\rangle \leftrightarrow }|\mathbf{2},2\mathbf{\rangle }
$ and $|\mathbf{1},6\mathbf{\rangle \leftrightarrow }|\mathbf{3},4\mathbf{%
\rangle }$; we
verified that for certain values of parameters this fact persists to the
fourth order in $g_{0}/\Delta _{-}$ as well.

\section{Numeric results}

\label{sec3}

\begin{figure}[t]
\begin{center}
\includegraphics[width=0.48\textwidth]{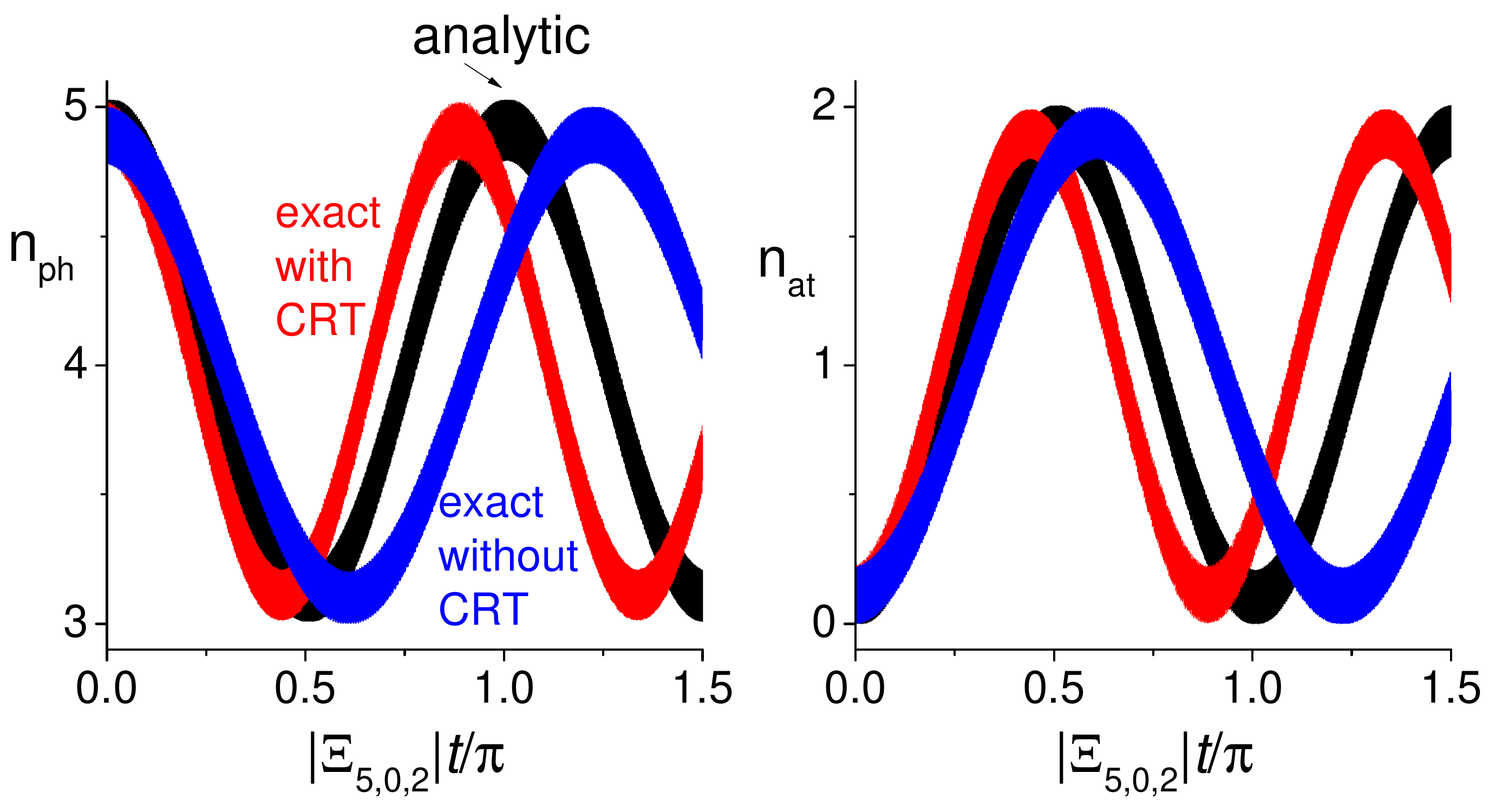} {}
\end{center}
\caption{(Color online) Comparison of the approximate analytic and exact
numeric results (with and without CRT) for $N=2$. The initial state is $|%
\mathbf{0}\rangle \otimes |5\rangle $. $n_{ph}$ and $n_{at}$ stand for the
average numbers of photons and atomic excitations, respectively. }
\label{F1}
\end{figure}
\begin{figure}[h]
\begin{center}
\includegraphics[width=0.48\textwidth]{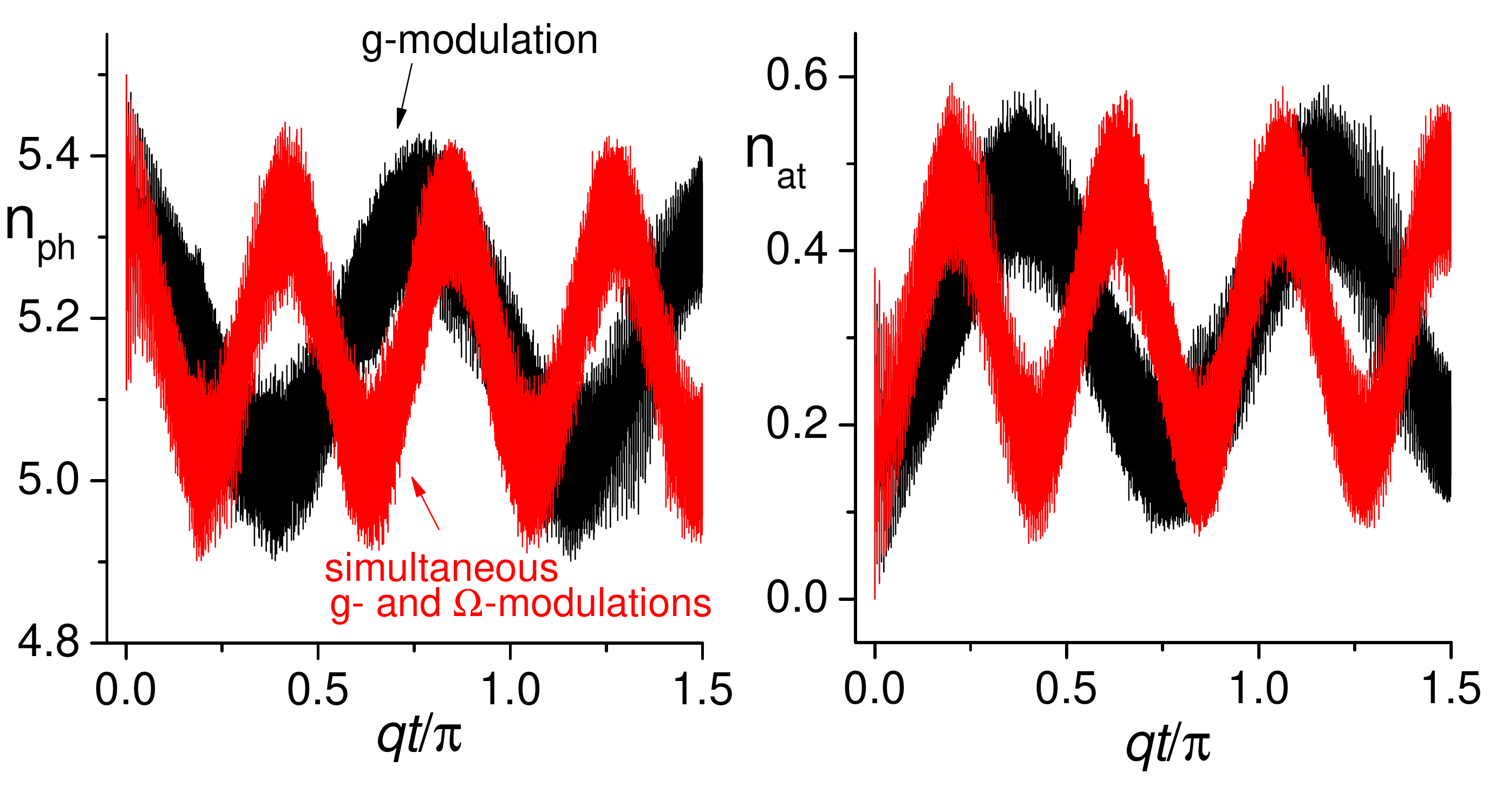} {}
\end{center}
\caption{(Color online) Exact numeric dynamics of the average excitation
numbers under the sole $g$-modulation (black) and the simultaneous
modulation of $g$ and $\Omega $ (red). The initial state is $|\mathbf{0}%
\rangle \otimes |\protect\alpha \rangle $, $\protect\alpha =5.5$ and $N=6$. }
\label{F2}
\end{figure}

To check our analytic predictions we solved numerically the Schr\"{o}dinger
equation for the original Hamiltonian (\ref{H1}). In Fig. \ref{F1} we
compare the exact numeric results, with and without CRT, to the approximate
formulas (\ref{dor}) -- (\ref{gor}). We plot the average number of photons $%
n_{ph}=\langle \psi |\hat{n}|\psi \rangle $ and the average number of atomic
excitations $n_{at}=\langle \psi |\sum_{k=1}^{N}k\sigma _{k,k}|\psi \rangle $
for the initial state $|\mathbf{0}\rangle \otimes |5\rangle $ and parameters
$N=2$, $g_{0}\sqrt{N}/\omega _{0}=8\times 10^{-2}$, $\Delta _{-}=-9g_{0}%
\sqrt{N}$, $\varepsilon _{g}/g_{0}=10^{-1}$, $\phi _{g}=0$, $\varepsilon
_{\Omega }=\varepsilon _{\omega }=0$. As expected, the resonant modulation
frequencies vary depending on whether the CRT are taken into account or not:
$\eta _{r}=2|\Delta _{-}|\times 1.0678$ with CRT and $\eta _{r}=2|\Delta
_{-}|\times 1.0540$ without CRT. The analytic and numeric results agree
qualitatively, though there is a roughly $20\%$ difference in the analytic
and actual transition rates $|\Xi _{5,0,2}|$. Such discrepancy is not
surprising, since for the above parameters the required inequalities $\sqrt{%
2n}g_{0}/\Delta _{-}\ll 1$ and $|\Delta _{-}|\ll \omega _{0}$ are only
barely satisfied \cite{foot1}. The apparent broad width of the curves is
explained by fast low-amplitude oscillations due to the off-resonant photon
exchange inherent to the dispersive regime, as inferred from the
wavefunction (\ref{psit}) and the expression for the dressed states (\ref%
{dress}).

In Fig. \ref{F2} we consider a more realistic initial state $|\mathbf{0}%
\rangle \otimes |\alpha \rangle $, where $|\alpha \rangle =e^{-|\alpha
|^{2}/2}\sum_{n=0}^{\infty }\alpha ^{n}/\sqrt{n!}$ stands for the cavity
\emph{coherent state} with $\alpha =\sqrt{5.5}$ [so that the initial
probability of 5 photons is $P_{ph}(5)\approx 0.17$]. For the sake of
compactness we only present the exact numeric results in the presence of
CRT. We set $N=6$ and consider the $g$-modulation (with parameters $g_{0}%
\sqrt{N}/\omega _{0}=8\times 10^{-2}$, $\Delta _{-}=-9g_{0}\sqrt{N}$, $%
\varepsilon _{g}/g_{0}=10^{-1}$, $\phi _{g}=0$ and $\eta _{r}=2|\Delta
_{-}|\times 1.0389$), as well as the simultaneous modulation of $g$ and $%
\Omega $ with the additional parameters $\varepsilon _{\Omega }/|\Delta
_{-}|=10^{-1}$ and $\phi _{\Omega }=\pi $ (in this case $\eta _{r}=2|\Delta
_{-}|\times 1.0388$). These modulation frequencies were adjusted to promote
the transition $|\mathbf{0},5\rangle \leftrightarrow |\mathbf{2},3\rangle $,
and we defined $q\equiv |\Xi _{5,0,2}(\varepsilon _{\Omega }=\varepsilon
_{\omega }=0)|$ (i. e., $q$ is the transition rate under the pure $g$%
-modulation). We observe that excitations are transferred between the cavity
and far-detuned atoms, and under the simultaneous $g$- and $\Omega $%
-modulations the oscillations are roughly twice faster than under the sole $%
g $-modulation, in agreement with Eq. (\ref{rate}).

\begin{figure}[t]
\begin{center}
\includegraphics[width=0.48\textwidth]{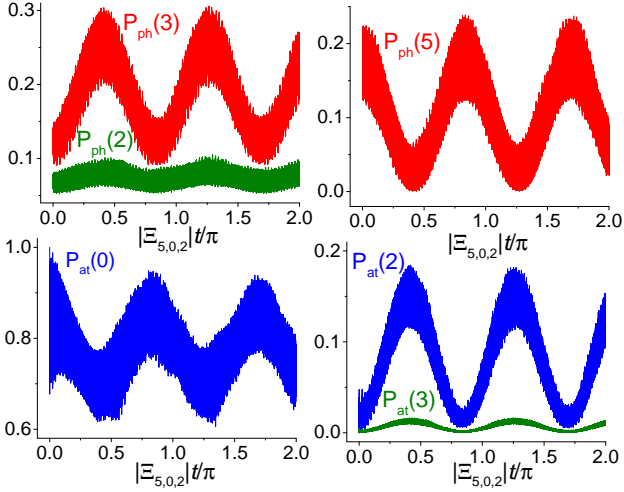} {}
\end{center}
\caption{(Color online) Exact numeric dynamics of probabilities $P_{ph}(k)$
and $P_{at}(m)$ for $N=6$ and the simultaneous modulation of $g$ and $\Omega
$. The initial state is $|\mathbf{0}\rangle \otimes |\protect\alpha \rangle $%
, $\protect\alpha =\protect\sqrt{5.5}$, and the modulation induces the
transition $|\mathbf{0},5\rangle \rightarrow |\mathbf{2},3\rangle $. }
\label{F3}
\end{figure}

To attest that the periodic behavior of $n_{ph}$ and $n_{at}$ indeed
corresponds to a two-photon exchange, in Fig. \ref{F3} we plot the
probability $P_{ph}(k)$ of $k$ photons and the probability $P_{at}(m)$ of $m$
atomic excitations under the simultaneous $g$- and $\Omega $-modulations
(discussed in Fig. \ref{F2}). We observe that the main transition occurs
between the states $|\mathbf{0},5\rangle \ $and $|\mathbf{2},3\rangle $,
although there are unwanted couplings between other states owing to the
off-resonant one-photon exchange. As an example, we illustrate small
oscillations between the states $|\mathbf{2},3\rangle \leftrightarrow |%
\mathbf{3},2\rangle $, inferred from the periodic oscillation of
probabilities $P_{ph}(2)$ and $P_{at}(3)$ at the same rate as the
probabilities $P_{ph}(3)$ and $P_{at}(2)$.

\begin{figure}[t]
\begin{center}
\includegraphics[width=0.48\textwidth]{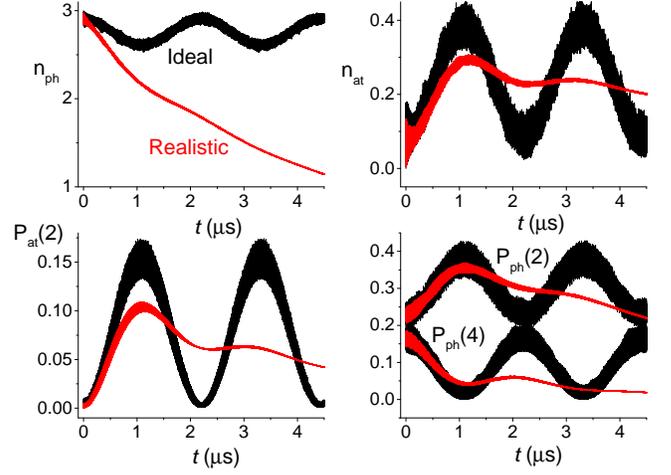} {}
\end{center}
\caption{(Color online) Comparison of the dynamics under realistic and ideal
conditions in two-qubit circuit QED architecture. Here we consider the $g$%
-modulations of both qubits, $\protect\omega _{0}/2\protect\pi =10\,$GHz and
the initial state $|g^{(1)},g^{(2)}\rangle \otimes |\protect\alpha \rangle $%
, $\protect\alpha =\protect\sqrt{3}$. The modulation drives the transition $|\mathbf{0}%
,4\rangle \leftrightarrow |\mathbf{2},2\rangle $.}
\label{F4}
\end{figure}

\subsection{Simulation under realistic conditions}\label{seccc}

The above numeric results apply to an ideal situation, namely, strictly
identical atoms and dissipation-free environment. To asses the experimental
feasibility of our proposal in circuit QED, we consider a realistic scenario
of two slightly different artificial atoms coupled to a single-mode
waveguide resonator under weak Markovian dissipation. The dynamics is now
governed by the master equation%
\begin{equation}
d\hat{\rho}/dt=\frac{1}{i\hbar }[\tilde{H},\hat{\rho}]+\mathcal{\hat{L}}\hat{%
\rho}  \label{mer}
\end{equation}%
\begin{equation*}
\tilde{H}/\hbar =\omega \hat{n}+\sum_{l=1}^{2}\left[ \frac{\Omega ^{(l)}}{2}%
\hat{\sigma}_{z}^{(l)}+g^{(l)}(\hat{a}+\hat{a}^{\dagger })(\hat{\sigma}%
_{+}^{(l)}+\hat{\sigma}_{-}^{(l)})\right] ~,
\end{equation*}%
where $\hat{\rho}$ is the total density operator and $\mathcal{\hat{L}}$ is
the Liouvillian. To get a rough estimative we solved numerically the
\lq standard\rq\ phenomenological master
equation \cite{bla} for zero-temperature reservoirs \cite{foot2} 
\begin{equation}
\mathcal{\hat{L}}\hat{\rho}=\kappa \mathcal{D}[\hat{a}]\hat{\rho}%
+\sum_{l=1}^{2}\left( \gamma ^{(l)}\mathcal{D}[\hat{\sigma}_{-}^{(l)}]+\frac{%
\gamma _{\phi }^{(l)}}{2}\mathcal{D}[\hat{\sigma}_{z}^{(l)}]\right) \hat{\rho%
}\,,
\end{equation}%
where $\mathcal{D}[\hat{O}]\hat{\rho}\equiv (2\hat{O}\hat{\rho}\hat{O}%
^{\dagger }-\hat{O}^{\dagger }\hat{O}\hat{\rho}-\hat{\rho}\hat{O}^{\dagger }%
\hat{O})/2$ is the Lindbladian superoperator. The constant parameters $%
\kappa $, $\gamma ^{(l)}$ and $\gamma _{\phi }^{(l)}$ denote the cavity
damping and the $l$-th qubit's relaxation and pure dephasing rates,
respectively.

In Fig. \ref{F4} we compare the dynamics for the ideal and realistic
scenarios under the $g$-modulations and the initial cavity coherent state $%
|g^{(1)},g^{(2)}\rangle \otimes |\alpha \rangle $, where $\alpha =\sqrt{3}$.
For the realistic case we set: $g^{(l)}=g_{0}^{(l)}+\varepsilon _{g}^{(l)}\sin
(\eta t)$, $g_{0}^{(1)}/\omega _{0}=5.66\times 10^{-2}$, $%
g_{0}^{(2)}=1.01g_{0}^{(1)}$, $\varepsilon _{g}^{(l)}/g_{0}^{(l)}=0.1$, $%
\Delta _{-}^{(1)}\equiv \omega _{0}-\Omega ^{(1)}=-0.72\omega _{0}$, $\Delta
_{-}^{(2)}=1.02\Delta _{-}^{(1)}$, $\kappa /g_{0}^{(1)}=\gamma
^{(l)}/g_{0}^{(l)}=5\times 10^{-5}$, $\gamma _{\phi }^{(l)}=\gamma ^{(l)}$
and $\eta _{r}=2|\Delta _{-}^{(1)}|\times 1.0632$. For the ideal case
$g_{0}/\omega _{0}=5.66\times 10^{-2}$, $\varepsilon _{g}/g_{0}=0.1$, $%
\Delta _{-}=-0.72\omega _{0}$ and $\eta _{r}=2|\Delta _{-}|\times 1.0531$
(the modulation frequencies were chosen to induce the transition $|\mathbf{0}%
,4\rangle \leftrightarrow |\mathbf{2},2\rangle $). Such parameters are
compatible with the current circuit QED architectures, where typically $%
\omega _{0}/2\pi =10\,$GHz \cite{paik,ger,3at0,3at1,3at2,saito,reagor,ofec,schuster}. We see that for
initial times, $t\lesssim 1\,\mu $s, the two-photon exchange can be proved
via measurements of the average number of atomic excitations $n_{at}$ or the
probabilities $P_{at}(2)$, $P_{ph}(2)$ and $P_{ph}(4)$, whereas the
measurement of the average photon number is of little help due to
overwhelming effects of dissipation.

\section{Conclusions}

\label{sec4}

We showed analytically and numerically that effective two-photon exchange
interaction between a single cavity mode and $N>1$ off-resonant qubits can
be achieved by externally modulating any system parameter at frequency $\eta
\approx 2|\Delta _{-}|$, where $\Delta _{-}$ is the average atom--field
detuning. This effect originates from the \lq
rotating\rq\ terms in the Dicke (or Tavis-Cummings)
Hamiltonian, but the associated transition rate is quite small due to the
multiplicative factor $(g_{0}/\Delta _{-})^{3}$. Closed analytical
description was derived under the assumption of weak atom--field coupling,
and a good agreement with exact numeric results was observed even for
moderate coupling strengths. For a simultaneous modulation of different
parameters the transition rate can be increased by properly adjusting the
initial phases. Regarding the experimental feasibility, we demonstrated that
for $N=2$ our proposal can be implemented in the current circuit QED
architectures on the timescales $\sim 1\,\mu $s, which could be further reduced through
an increase in the modulation amplitudes, atom--field coupling strength or
the number of qubits.

\begin{acknowledgments}
The author acknowledges a partial support of the Brazilian agency CNPq
(Conselho Nacional de Desenvolvimento Cient\'{\i}fico e Tecnol\'{o}gico).
\end{acknowledgments}


\begin{thebibliography}{99}
\bibitem{foot1}The difference between the numeric transition rates with and without CRT is
roughly $30\%$ because for our parameters, $|\Delta _{-}|/\omega _{0}=0.72$,
the CRT are not negligible.
\end{thebibliography}
\end{document}